\documentclass[pdflatex,sn-mathphys-num]{sn-jnl}

\usepackage{array}
\usepackage{tabularx}
\usepackage{booktabs}
\usepackage{ragged2e}
\usepackage{float}
\usepackage{placeins}
\usepackage{longtable}
\usepackage{booktabs}
\usepackage{caption}
\usepackage[none]{hyphenat}

\usepackage{graphicx}%
\usepackage{multirow}%
\usepackage{amsmath,amssymb,amsfonts}%
\usepackage{amsthm}%
\usepackage{mathrsfs}%
\usepackage[title]{appendix}%
\usepackage{xcolor}%
\usepackage{textcomp}%
\usepackage{manyfoot}%
\usepackage{booktabs}%
\usepackage{algorithm}%
\usepackage{algorithmicx}%
\usepackage{algpseudocode}%
\usepackage{listings}%

\theoremstyle{thmstyleone}%
\theoremstyle{thmstyletwo}%

\theoremstyle{thmstylethree}%

\begin{document}

\title[Article Title]{A Robust and Explainable Transformer-Based Framework for Phishing Email Detection}


\author*[1]{\fnm{Sajad} \sur{U P}}\email{sajadup2@gmail.com}

\affil*[1]{\orgname{Independent Researcher}, \orgaddress{\country{India}}}

\abstract{Phishing and related cyber threats are becoming increasingly sophisticated, with email-based phishing remaining the most persistent attack vector. These attacks exploit human vulnerabilities to deliver malware or gain unauthorized access to sensitive information. Transformer-based models enhance phishing detection through robust contextual language understanding; yet they are often regarded as black boxes due to a lack of interpretability. Moreover, recent AI-enabled attacks further undermine model resilience. To address these challenges, this work proposes a lightweight phishing detection framework based on DistilBERT, a lightweight Transformer model. Robustness to embedding-level perturbations and character-level input noise is enhanced through gradient-based adversarial training using the Fast Gradient Method (FGM), combined with stochastic character-level perturbations. To improve transparency, three prominent Explainable AI (XAI) methods, LIME (Local Interpretable Model-agnostic Explanations), SHAP (SHapley Additive exPlanations), and IG (Integrated Gradients), are integrated to interpret model decision-making. A structured rule-based prompt combines model predictions and XAI features to guide Flan-T5-Small in generating plain-language, evidence-based explanations. Experimental results demonstrate that the proposed framework outperforms a standard DistilBERT-based detection model trained without robustness enhancements in terms of accuracy and resilience. This integrated approach helps bridge the gap between model reliability and user trust, advancing transparent phishing detection.}

\keywords{Phishing detection, DistilBERT, Adversarial training, Explainable AI, Flan-T5, Natural language explanations}



\maketitle

\section{Introduction}\label{sec1}

As cybercrime grows in scale and sophistication, AI-enhanced phishing and social engineering have emerged as major global cybersecurity threats \cite{wef2025cybersecurity}. Among these, phishing attacks that deceive users by impersonating trusted sources remain the most prevalent vector, particularly through email, due to the widespread availability of public email accounts. Attackers often design phishing emails to closely mimic trusted sources, posing significant challenges to both users and traditional filtering systems in distinguishing them from legitimate communications. From an economic perspective, large-scale phishing campaigns have low operating costs due to their minimal infrastructure requirements. Despite repeated awareness campaigns, the click-through rate for phishing emails remains high. This can lead users to download malicious attachments, disclose credentials, or click harmful links. Phishing can lead to data breaches, reputational damage, and legal repercussions.

The recent surge of publicly available Large Language Models (LLMs) presents a cybersecurity dilemma \cite{ferrag2025genai_cybersecurity}. While they enhance productivity, they also enable highly realistic phishing attacks. Moreover, adversarial attacks use slight text modifications to mislead models, bypassing standard phishing detection mechanisms \cite{gholampour2023adversarial_phishing}.  Deep learning (DL) approaches, particularly Transformer architectures \cite{vaswani2017attention}, have gained increasing attention for their ability to capture complex linguistic patterns. Among them, BERT-style Transformer models effectively capture contextual information through bidirectional self-attention over the entire text sequence \cite{devlin2019bert}. Despite their strong performance, Transformer-based models remain sensitive to adversarial perturbations and often function as black-box systems. This lack of robustness and transparency limits user trust and real-world applicability in digital security settings. 

In this work, DistilBERT (a distilled version of BERT) is employed for phishing email detection \cite{sanh2020distilbert}. By utilizing approximately 66 million parameters, it balances computational efficiency with high performance. To enhance security, the model is integrated with a single-step adversarial training strategy using FGM \cite{gong2018adversarial_texts, liu2024telecom_fraud_advtrain}. This approach applies gradient-based perturbations to the embedding layer, enabling the model to maintain stable representations under manipulation. Furthermore, stochastic character-level perturbations (insertion, deletion, and substitution) are used as data augmentation to simulate real-world evasion. Consequently, the model becomes resilient to both embedding-level attacks and character-level obfuscation.

In addition, this work integrates three local XAI techniques within the phishing detection pipeline: LIME \cite{ribeiro2016lime}, SHAP \cite{lundberg2017shap}, and IG \cite{sundararajan2017ig}. These techniques quantify feature importance by assigning attribution weights to specific tokens, highlighting their influence on the model’s detection decisions. For plain-language explanations, the prediction, confidence score, and top influential feature tokens extracted by the LIME model are converted into concise textual summaries using clear, rule-based prompts. These prompts guide the Flan-T5-Small model, a lightweight instruction-tuned sequence-to-sequence transformer from Hugging Face, to generate user-friendly security explanations \cite{chung2024instruction_finetuning}. This layered approach helps users understand why an email was flagged and supports informed decision-making regarding subsequent actions. This research aims to contribute to ongoing studies and practical efforts toward enhancing the implementation of existing phishing email detection systems. 

\vspace{8pt}
The major contributions of this research are summarized as follows:
\begin{itemize}
  \item Enhanced Hybrid Robustness: This work demonstrates that a hybrid robustness strategy combining FGM-based embedding-level adversarial training with character-level noise augmentation significantly improves DistilBERT’s resilience to character-level perturbation attacks. This approach provides a computationally efficient alternative to multi-step adversarial defenses, making it suitable for real-time email security applications.
  \vspace{6pt}
  \item Comparative XAI Evaluation: A systematic comparison of LIME, SHAP, and IG is conducted using character-level perturbations and selective token removal on robustly trained models. This analysis quantifies the interpretability and qualitative robustness of attribution methods under realistic input distortions, offering a benchmark for transparent model explanations.
    \vspace{6pt}
  \item Evidence-Grounded Natural Language Explanations: The study establishes an explanation layer using Flan-T5-Small to convert feature attributions and prediction results into concise, human-readable narratives. Unlike static alerts, these explanations are directly grounded in model evidence to support informed security decision-making.
    \vspace{6pt}
  \item An End-to-End Robust Interpretable Lightweight Security Framework: By unifying robust training mechanisms, XAI, and natural language rationalization, this research provides a deployable pipeline that addresses the gap between model robustness and user trust, advancing the state of transparent phishing detection.
      \vspace{6pt}
\end{itemize}

The rest of the paper is organized as follows. Section 2 reviews related work and provides an overview of classic and modern phishing detection frameworks, along with core studies in cybersecurity, model robustness, and XAI. Section 3 describes the proposed framework methodology in detail. Section 4 presents the experimental results. Section 5 provides an extensive discussion of the results, along with an analysis of limitations and ethical considerations. Section 6 concludes the paper and highlights potential directions for future work.
 
\section{Related work}\label{sec2}
\subsection{Evolution of Phishing Detection} 
Phishing attacks have evolved alongside advances in Internet technologies. This increased sophistication has motivated the development of diverse detection and prevention techniques. Early phishing and spam mitigation approaches largely relied on static blacklists and rule-based filtering systems, which offered little flexibility in response to evolving malicious tactics \cite{khonji2013phishing_survey}. The transition toward machine learning-based approaches initiated a notable transformation, enabling systems to identify patterns and anomalies in email content through data-driven techniques. One of the earliest comprehensive evaluations of the Naive Bayes classifier for anti-spam filtering was presented in \cite{androutsopoulos2000spam}. This work conducted a pioneering experimental comparison between Naive Bayesian and keyword-based anti-spam filtering techniques.  

Deep learning has emerged as a promising machine learning approach for phishing detection in recent years, marking a significant advancement in detection capabilities. Several deep learning applications utilized architectures such as Convolutional Neural Networks (CNN) and Long Short-Term Memory (LSTM) networks, demonstrating significant performance gains over traditional machine learning \cite{alshingiti2023cnn_lstm_phishing, altwaijry2024comparative_phishing}. A flexible and in-depth phishing email detection system, referred to as D-Fence, was introduced in \cite{lee2021dfence}. This system employs a multi-modular architecture with specialized learning components for structure, textual content, and URL analysis, which collectively enable effective phishing detection. Ensemble-based approaches combining heterogeneous algorithms and classifiers achieved competitive accuracy but often relied on complex feature pipelines that may reduce real deployment feasibility \cite{sa2024ensemble_phishing}. 

A systematic review of deep learning techniques for phishing email detection is presented in \cite{kyaw2024systematic_phishing_dl, do2022taxonomy_phishing}. These reviews identified critical research gaps, including the need for models that adapt to novel phishing behaviors. Comprehensive anatomical studies of phishing attacks have provided detailed insights into attack vectors, social engineering tactics, and evolving strategies employed by adversaries \cite{alkhalil2021phishing_anatomy}. Despite these gains, research consistently highlights that the adaptive nature of phishing requires models that move beyond simple accuracy toward higher resilience and transparency \cite{okafor2024dl_cybersecurity}.

\subsection{Transformer-Based Deep Learning Architectures} 
The introduction of Transformer architectures \cite{vaswani2017attention} revolutionized natural language processing tasks, including phishing email detection. A pre-trained language model based on the transformer architecture with a bidirectional attention mechanism, BERT, was introduced in \cite{devlin2019bert}. Unlike traditional left-to-right or right-to-left language models, this approach captures contextual information from both directions simultaneously. To address the high computational cost associated with large-scale models such as BERT, a distilled variant DistilBERT is introduced \cite{sanh2020distilbert}, \cite{huggingface_distilbert_docs}. This lightweight model retains approximately 97\% of the original language understanding capability while reducing model size by about 40\% and achieving nearly 60\% faster inference, making it more suitable for real-time deployment. 

\cite{otieno2023bert_phishing} presents transformer-based classification results for phishing and non-phishing emails using the BERT language model, demonstrating the effectiveness of BERT for phishing email detection. BERT-based models, including DistilBERT and RoBERTa, have been fine-tuned for phishing detection, achieving strong performance \cite{songailaite2023bert_phishing, jamal2024improved_transformer_phishing}. Comparative investigations highlight the superior performance of Transformer architectures like DistilBERT, BERT, XLNet, ALBERT, and RoBERTa over traditional models like logistic regression and random forests in phishing email tasks \cite{melendez2024ml_vs_transformers}. \cite{asliyuksek2025multimodal_distilbert} proposed a multimodal spam email classification approach that combines DistilBERT-based textual representations with structural email features, demonstrating improved performance. Subsequent studies extended transformer-based phishing detection by integrating additional architectural components. A unified deep learning model that integrates BERT embeddings, multi-head attention, and GRU networks is proposed to improve phishing email detection performance \cite{hosseinzadeh2025adaptive_optimization}. \cite{safran2025phishinggnn} introduces PhishingGNN, a phishing email detection framework that combines transformer-based feature extraction with graph attention networks to capture relational and contextual patterns in emails.

\subsection{Adversarial Robustness and Training}
Adversarial training has emerged as a critical strategy for enhancing the robustness of AI-based models. In the context of email security, an analysis of the adversarial robustness of phishing email detection models presented in \cite{gholampour2023adversarial_phishing}, shows that even high-performing deep learning classifiers are highly susceptible to minor textual perturbations. Foundational work on fast gradient-based adversarial example generation, initially developed for image-based models \cite{goodfellow2015adversarial}, has significantly influenced the adoption of adversarial training strategies in the text domain. Subsequent surveys and case studies have categorized gradient-based methods for the generation of adversarial text samples \cite{liu2024telecom_fraud_advtrain}, \cite{alsmadi2022adversarial_text_survey}. \cite{liu2023adv_training_text} proposes an adversarial training method, HNN-GRAT, which uses RoBERTa for feature extraction and applies a gradient reversal layer to fuse original and inverted gradients, improving robustness in text classification. Studies further indicate that such training not only improves resilience against specific evasion tactics but also enhances the generalizability of language models across diverse attack types \cite{altinisik2023impact_adv_training}. Collectively, prior work indicates that integrating adversarial training is essential for robust environments, including phishing detectors \cite{goodfellow2015adversarial, alsmadi2022adversarial_text_survey, liu2023adv_training_text, altinisik2023impact_adv_training}.

\subsection{Interpretability in Cybersecurity} 
A growing focus has been placed on explainability and interpretability techniques to improve transparency in the cybersecurity domain. XAI has emerged as a critical requirement for security-oriented deep learning systems, where trust, accountability, and human oversight are essential. A detailed literature survey on XAI for cybersecurity is presented in \cite{charmet2022xai_cyber_survey}, establishing both theoretical foundations and practical requirements for applying XAI in security-related applications. These papers discuss how explainability techniques can enhance threat detection and response capabilities across multiple security domains. Domain-specific analyses of XAI methods have demonstrated their applicability to malware detection, highlighting both interpretability benefits and a lack of generalization in evaluating XAI models \cite{mohamed2025ai_cybersecurity_review}. A web-based AI platform for phishing email detection is proposed with a strong emphasis on the interpretability of model predictions \cite{subaiey2024robust_xai_platform}, where emails are classified into distinct categories based on their textual patterns.

Interpretability in cybersecurity has been widely addressed through post-hoc explanation techniques such as LIME, SHAP, and IG to improve transparency and trust in deep learning based security systems \cite{ribeiro2016lime, lundberg2017shap, sundararajan2017ig}. Review and survey studies have identified LIME and SHAP as commonly used model-agnostic methods for explaining predictions in cybersecurity applications, including phishing and malware detection, by highlighting influential input features or tokens \cite{saqib2024xai_malware, capuano2022xai_cyber, reynaud2025review_xai_cyber}. Gradient-based attribution methods, particularly IG, provide stable and theoretically grounded explanations for neural text classifiers and have been effectively applied to analyze model behavior under adversarial conditions \cite{moraliyage2025ig_adversarial_text}.

\subsection{Research Gaps and Motivations} 
Despite notable advances in phishing detection, classification, robustness, and explainability, existing studies typically address these aspects in isolation. Most models primarily optimize classification accuracy. In contrast, robustness-oriented training approaches improve resilience but often lack interpretability, thereby limiting user trust and real-world applicability. Similarly, XAI-focused studies often restrict explanations to technical feature attributions that are difficult for non-expert users to interpret and do not directly support informed decision-making.

These gaps highlight the need for an integrated approach that simultaneously addresses robustness and explainability in a phishing detection system. Motivated by this limitation, this study proposes a unified, lightweight, end-to-end phishing detection framework that jointly incorporates robustness-enhancing training, comparative XAI analysis, and language-model-driven plain-language explanations. By combining robustly trained transformer models with both technical and human-readable explanations, the proposed approach aims to improve reliability, transparency, and user trust in real-world phishing detection systems.

\subsection{Comparison with Prior Work} 
Table 1 compares the proposed framework (Explainable and Robust DistilBERT) with representative prior studies across four key dimensions: Adversarial/ Robustness Analysis, Explainability, Comparative XAI Evaluation, and User-Level Interpretability.

\FloatBarrier
\begin{table}[h]
\caption{Comparison of the Proposed Framework with Prior Work}\label{tab:comparison_prior_work}
\begin{tabular*}{\textwidth}{@{\extracolsep\fill}llcccc}
\toprule
Prior Work & Domain & Adv./Robustness & Explainability & XAI Eval. & User Interpret. \\
\midrule
\cite{gholampour2023adversarial_phishing} & Phishing      & \checkmark & $\times$   & $\times$   & $\times$   \\
\cite{gong2018adversarial_texts}          & Text Class.   & \checkmark & $\times$   & $\times$   & $\times$   \\
\cite{liu2023adv_training_text}           & Text Class.   & \checkmark & $\times$   & $\times$   & $\times$   \\
\cite{subaiey2024robust_xai_platform}     & Phishing      & \checkmark & \checkmark & $\times$   & Limited    \\
\cite{moraliyage2025ig_adversarial_text}  & Text Class.   & \checkmark & \checkmark & $\times$   & $\times$   \\
\textbf{Proposed}                         & \textbf{Phishing} & \checkmark & \checkmark & \checkmark & \checkmark \\
\botrule
\end{tabular*}
\footnotetext{Note: \checkmark~= present; $\times$ = absent; ``Limited'' = partial support. Adv.~= Adversarial; Class.~= Classification; XAI~= Explainable AI; Interpret.~= Interpretability.}
\end{table}
\FloatBarrier

\section{Methodology}\label{sec3}
\subsection{Data Preparation and Preprocessing}
\subsubsection{Dataset Description}
The experiments use a publicly available phishing email dataset from Hugging Face (ID: zefang-liu/phishing-email-dataset) \cite{liu2024phishing_dataset}, which contains approximately 18,650 email samples. The class distribution is imbalanced, with 60.71\% legitimate and 39.29\% phishing emails. Each record contains two main fields: Email Text and Email Type, where Email Type is encoded as 0 (Legitimate) and 1 (Phishing) for binary classification.

\vspace{6pt}
\subsubsection{Data Cleaning and Text Normalization}
All email content is converted to lowercase, and extra whitespace is removed to ensure textual consistency. This normalization step improves robustness and generalization of the DistilBERT-based phishing detection model.

\vspace{6pt}
\subsubsection{Tokenization and Encoding}
The preprocessed email text is tokenized using DistilBertTokenizerFast, a high-performance, Rust-based tokenizer from the Hugging Face Transformers library. The tokenizer converts text into subword tokens and corresponding input IDs compatible with the DistilBERT model, enabling efficient and consistent input encoding for phishing detection. Since DistilBERT uses WordPiece subword tokenization, words may be split into multiple subword fragments based on the learned vocabulary. 

\vspace{6pt}
\subsubsection{Class Imbalance Handling and Splitting}
Stratified splitting was used to preserve the original class distribution across the training (70\%), validation (15\%), and test (15\%) sets. To further address class imbalance during training, a class-weighted cross-entropy loss was applied, with weights computed from the training data using a balanced weighting scheme.

\subsection{Framework Architecture}
The proposed framework comprises two main components: a robustly trained DistilBERT detection module and a user-centric plain-language explanation module. After data acquisition, email contents undergo preprocessing and tokenization, and the resulting token sequences are used to form a unified input representation for the transformer encoder. The encoder produces contextualized text features, which are passed to a classification layer to generate class logits. A softmax function converts these logits into class probabilities, yielding the final prediction (phishing or legitimate) along with a confidence score. This output provides both the model decision and the necessary inputs for the subsequent explanation stage.
The training process is reinforced with robustness-enhancing strategies, including FGM-based adversarial perturbation and character-level noise augmentation, to improve resilience against small, malicious text modifications. 

\begin{figure}[ht]
  \centering
  \includegraphics[width=\columnwidth]{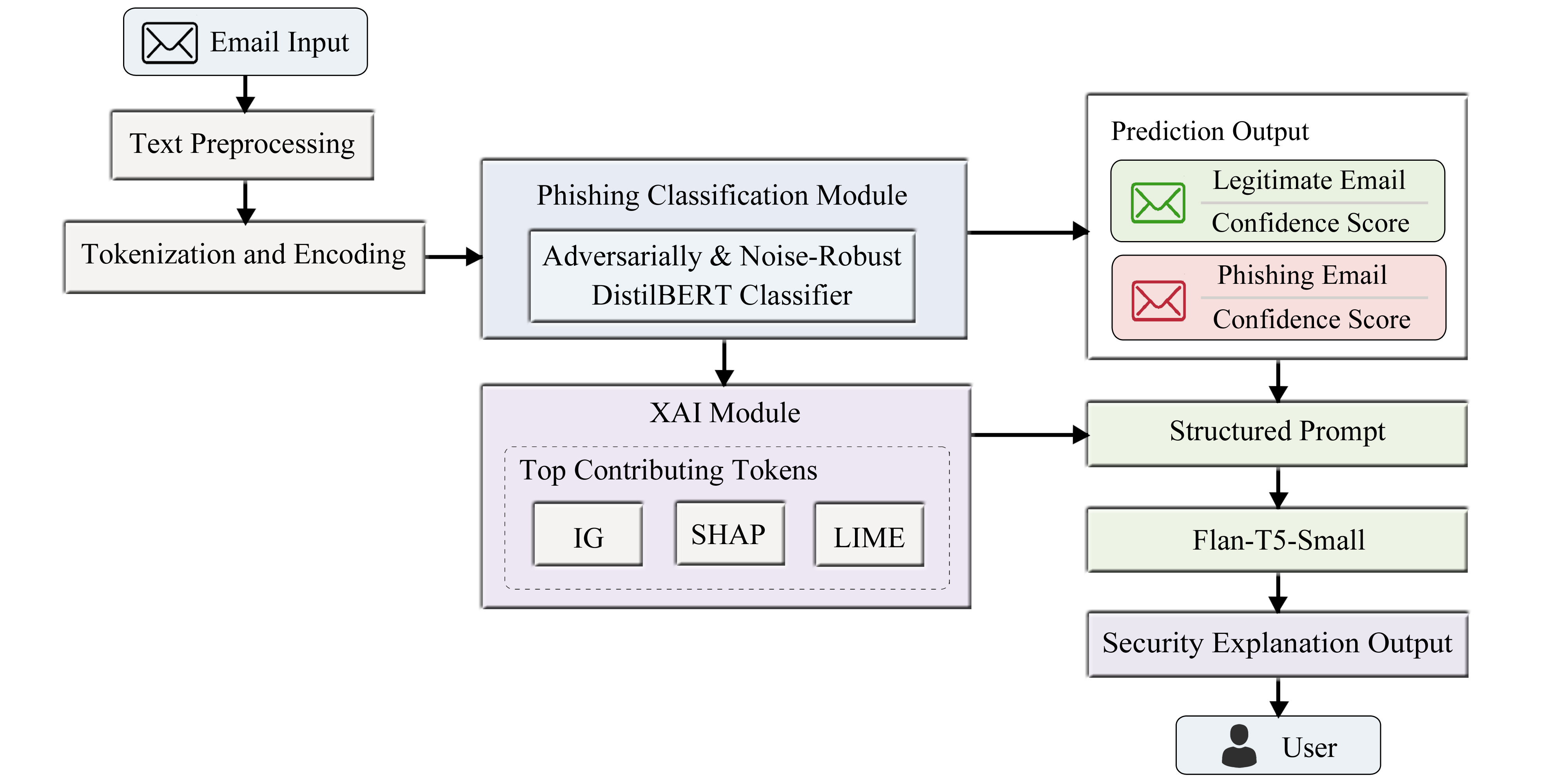}
  \caption{End-to-end pipeline from email input to security explanation output}
  \label{fig:workflow_diagram}
\end{figure}

Following classification, LIME, SHAP, and IG are applied to analyze token-level feature importance in the email text. For plain-language explanation generation, the most influential tokens identified by LIME, together with their importance scores and the model prediction, are mapped into a structured, tactic-specific prompt. This prompt is processed by an instruction-fine-tuned sequence-to-sequence model (Flan-T5-Small) using a Hugging Face Transformers pipeline to generate a cohesive, user-friendly security narrative. This unified workflow is illustrated in Figure ~\ref{fig:workflow_diagram}, which presents the end-to-end architecture of the proposed framework.

\FloatBarrier

\subsubsection{DistilBERT for Phishing Detection}
This research employs the pre-trained DistilBERT-base-uncased model as the backbone for phishing email detection. The model is instantiated using pre-trained weights from the Hugging Face Transformers library, together with the DistilBertTokenizerFast for efficient subword tokenization. Input emails are tokenized into sequences incorporating necessary special tokens, padding, and truncation. The input representation combines token embeddings for lexical information with positional embeddings to preserve word order. Consistent with its single-sequence design, DistilBERT omits segment embeddings. These representations are processed through stacked transformer layers employing multi-head scaled dot-product self-attention, enabling bidirectional contextual understanding and effective modeling of long-range dependencies across the email content. The final-layer hidden state corresponding to the [CLS] token is used as a fixed-length representation of the input sequence. This vector is processed by a task-specific pre-classifier and classification head, and the resulting logits are transformed into class probabilities via a softmax function to produce the final prediction and confidence score.

\subsubsection{Hybrid Robustness Training Using FGM and Character-Level Perturbations}
To improve model resilience, a hybrid robustness training strategy is adopted that combines embedding-level adversarial training using the FGM with stochastic character-level perturbations applied to input text during training. Character-level noise introduces discrete surface-form variations prior to tokenization, encouraging robustness to input-level manipulations that can disrupt subword segmentation. In contrast, FGM perturbs continuous embeddings after tokenization to penalize sensitivity to small gradient-based perturbations and improve local stability. By operating at different stages of the text processing pipeline, this hybrid strategy provides complementary protection. As a result, it improves resilience against both surface-level and embedding-level textual perturbations.

\vspace{6pt}
\textbf{Character-Level Noise Augmentation:}

Character-level noise is injected into a subset of training samples to simulate realistic evasion. Perturbations are applied to 30\% of training samples, with approximately 10\% of characters in each selected sample corrupted. These values balance robustness gains and semantic preservation, introducing variability without excessive distortion.
Corruption operations include random character insertion, deletion, and substitution, with substitutions biased toward visually similar characters (e.g., "o"→"0", "l"→"1", "a"→"@"). Perturbations are applied to raw text before tokenization, affecting both subword segmentation and contextual representations. Since these perturbations are not optimized to maximize model loss, character-level noise functions as data augmentation rather than adversarial attack generation.

\subsubsection{Embedding-Level Adversarial Training with FGM}
FGM-based adversarial training is applied directly to token embeddings during training. Unlike Fast Gradient Sign Method (FGSM), which perturbs inputs using only the sign of the gradient, FGM applies perturbations along the normalized gradient direction, yielding smoother perturbations in continuous embedding space.

The adversarial perturbation $\mathbf{\delta}$ is computed from the gradient of the clean loss with respect to the embedding representations:
\begin{equation}
\label{eq:fgm_perturb}
\mathbf{\delta} = \epsilon \cdot \frac{\nabla_{\mathbf{e}} \mathcal{L}_{\text{clean}}(\boldsymbol{\theta}, \mathbf{e}, y)}{\left\| \nabla_{\mathbf{e}} \mathcal{L}_{\text{clean}}(\boldsymbol{\theta}, \mathbf{e}, y) \right\|_2}
\end{equation}
where $\mathbf{e}$ denotes the token embedding matrix, $\mathcal{L}_{\text{clean}}$ is the class-weighted cross-entropy loss, $\boldsymbol{\theta}$ represents the model parameters, and $\|\cdot\|_2$ denotes the $\ell_2$ norm. The hyperparameter $\epsilon$ signifies the perturbation magnitude, which scales the gradient-based noise to ensure the adversarial embeddings remain within a controlled radius of the original representations. The perturbation magnitude is set to $\epsilon = 0.01$ based on preliminary tuning to enhance robustness while maintaining clean-data accuracy.

\subsubsection{Training Procedure}
For each training batch, the hybrid robustness training proceeds as follows:

\begin{itemize}
    \item \textbf{Character-level augmentation:} To simulate surface-level noise, 30\% of the training data is perturbed at the character-level, with about 10\% character corruption per sample.
    \item \textbf{Clean forward pass:} Compute the clean loss $\mathcal{L}_{\text{clean}}$ using original embeddings $\mathbf{e}$.
    \item \textbf{Adversarial perturbation:} Compute $\mathbf{\delta}$ via Eq. \ref{eq:fgm_perturb}. The perturbation is detached from the computational graph to prevent second-order gradient flow.
    \item \textbf{Adversarial forward pass:} Construct $\mathbf{e}_{\text{adv}} = \mathbf{e} + \mathbf{\delta}$ and compute $\mathcal{L}_{\text{adv}}$. where $\mathbf{e}_{\text{adv}}$ and $\mathcal{L}_{\text{adv}}$ denote the adversarial embeddings and corresponding loss.
\item \textbf{Joint Optimization:} The model parameters $\boldsymbol{\theta}$ are updated by minimizing a composite objective function:
    \begin{equation}
    \label{eq:total_loss}
    \mathcal{L}_{\text{total}} = \mathcal{L}_{\text{clean}} + \lambda \mathcal{L}_{\text{adv}}
    \end{equation}
    
    where $\lambda$ denotes the \textbf{adversarial weight coefficient} used to balance the trade-off between standard classification performance and adversarial resilience. This parameter is empirically set to $\lambda = 0.5$ to provide equal weighting to both loss components during the training phase.
\end{itemize}

\vspace{6pt}
\textbf{Selection of Adversarial Perturbation Strength ($\epsilon$), Adversarial Weight Coefficient ($\lambda$), and Learning Rate:}

To determine suitable hyperparameters for FGM-based adversarial training, an empirically guided tuning strategy was adopted using the Hugging Face phishing email dataset employed in this study, with the goal of identifying stable and effective configurations that balance robustness and clean-data performance. Validation behavior and overall training stability were used as the primary selection criteria. Based on these observations, the adversarial perturbation strength was set to $\epsilon = 0.01$, and the Adversarial Weight Coefficient was set to $\lambda = 0.5$. This combination provided a favorable trade-off between enhancing adversarial robustness and preserving classification accuracy on unperturbed inputs, and was therefore used consistently in all experiments. For model optimization, learning rates commonly reported as effective for DistilBERT fine-tuning were considered, and the final learning rate was set to $3 \times 10^{-5}$, as it yielded stable convergence and strong predictive performance on the validation set.
\vspace{6pt}
\par{Limitation:} Although these settings produced reliable results in the present experiments, the optimal values of $\epsilon$, $\lambda$, and the learning rate may vary across different scenarios. These hyperparameters are influenced by factors such as dataset diversity, sample size, class distribution, and noise characteristics. Consequently, when the framework is applied to different datasets or application domains, re-tuning may be required. It is therefore advisable to reassess model performance to ensure that changes in these parameter values do not adversely affect the results.

\subsubsection{XAI: Token-Level Feature Attribution and Explanation Stability}
To enhance transparency and evaluate the robustness of model explanations, three explainable attribution methods are employed: LIME, SHAP, and IG. These methods quantify the contribution of individual tokens to phishing classification decisions by assigning attribution scores that reflect each token’s influence on the predicted class. LIME explains individual predictions by locally approximating the classifier through input perturbations (token masking) and fitting an interpretable linear surrogate model. SHAP estimates approximate Shapley values grounded in cooperative game theory, measuring each token’s marginal contribution across different feature subsets. IG computes gradient-based attributions along a path from a baseline input (e.g., zero embeddings) to the actual input, satisfying axiomatic properties such as sensitivity and implementation invariance.

Standard implementations from widely used libraries were adopted to ensure reproducibility: LimeTextExplainer with 1{,}000 perturbation samples per explanation, Partition SHAP explainer with 500 model evaluations per sample, and Captum’s IG with 100 integration steps. Parameter values were chosen based on common defaults reported in prior work and preliminary experiments, providing a practical balance between attribution stability and computational efficiency, while accounting for differences in attribution mechanisms and computational complexity that lead to variations in sparsity, token overlap, and sensitivity to contextual dependencies across methods.

\vspace{6pt}
\textbf{Explanation Stability Evaluation:}

Beyond classification accuracy, the consistency of explanations under realistic input perturbations is evaluated through manual comparison. For selected test emails, the baseline DistilBERT and the proposed robust framework are first evaluated on the original email to obtain predictions, confidence scores, and top-10 token attributions from all three XAI methods. The same email is then manually perturbed using operations that resemble character-level noise augmentation, including character-level substitutions, deletions, insertions, and selective removal of influential tokens. Both models are re-evaluated on the perturbed version, and predictions, confidence scores, and token attributions are again recorded. 

Although no formal stability metric is computed, observable differences in confidence degradation and prediction shifts offer insight into the relative robustness of the baseline and proposed models. Changes in attribution patterns across perturbed inputs further highlight differences in explanation stability between the two approaches. Formal stability metrics are left for future work, as the goal here is comparative qualitative robustness under realistic phishing perturbations rather than absolute attribution invariance. Higher consistency between original and perturbed attributions suggests that the model relies on semantically meaningful indicators rather than fragile surface-form cues.

\subsubsection{Plain-Language Explanation via Flan-T5-Small}
To produce human-interpretable explanations, the framework employs Flan-T5-Small as a lightweight, post-hoc explanation generator. The model does not directly access classifier internals; instead, it receives a structured natural-language prompt constructed from classifier outputs and token-level attribution results. LIME-based token attributions are selected to condition the explanation generator. This decision is driven by LIME’s locality, model-agnostic nature, and, under the experimental setting, relatively focused attribution scores that align well with intuitive phishing indicators and are convenient for constructing concise input prompts for Flan-T5-Small. While SHAP and IG also produce reliable explanations, their attributions are often more distributed in these experiments, requiring additional selection heuristics to avoid overloading the prompt with excessive tokens. To maintain brevity and readability of the generated explanations, LIME is therefore prioritized for the downstream natural language generation stage.

\vspace{6pt}
The prompt design integrates the following components:

\begin{itemize}
    \item \textbf{Task Contextualization:} A directive requiring concise justification of the predicted class (phishing or legitimate).
    
    \item \textbf{Empirical Evidence Injection:} Inclusion of the predicted label, confidence score, and the top 10 most influential tokens ranked by absolute LIME attribution values.
    
    \item \textbf{Semantic Guidance:} Explicit instructions to relate highlighted tokens to common phishing heuristics, such as urgency cues, impersonation attempts, or credential harvesting.
    
    \item \textbf{Operational Constraints:} Instructions enforcing brevity, eliminating redundancy, and restricting reasoning to the provided evidence.
\end{itemize}

\vspace{6pt}
This text-to-text prompt design leverages Flan-T5’s instruction-following capabilities while reducing the risk of unsupported inferences. Given this input, Flan-T5-Small conditions on the provided evidence to map salient tokens to common phishing archetypes and generates a brief (1–2 sentence) human-readable justification. To ensure consistency and reproducibility in the generated explanations, greedy decoding is used during inference, resulting in deterministic outputs for identical inputs. However, variations in tokenization, library versions, or model checkpoints may still affect reproducibility across different computational environments, which is an important consideration for evaluation, user trust, and deployment in security-sensitive applications.

\vspace{2pt}
\subsection{Experimental Setup}

All experiments were conducted by fine-tuning DistilBERT (distilbert-base-uncased) on the zefang-liu/phishing-email-dataset using PyTorch and the Hugging Face Transformers library on Google Colaboratory with an NVIDIA Tesla T4 GPU. The model was instantiated using AutoTokenizer and AutoModelForSequenceClassification, ensuring consistency between tokenization and the pre-trained DistilBERT architecture. The maximum sequence length was set to 256 tokens. A batch size of 32 and a learning rate of $3 \times 10^{-5}$ were used for stable fine-tuning. Convergence was typically achieved within 3--6 epochs. For reproducibility, the random seed was fixed at 42 across all runs. To enhance robustness, FGM adversarial training was applied at the embedding-level with perturbation magnitude $\epsilon = 0.01$, and the adversarial loss weight was set to $\lambda = 0.5$, providing a trade-off between clean accuracy and adversarial robustness. In addition, character-level noise injection was applied as a data augmentation strategy during training: 30\% of the samples were randomly selected for perturbation, and within each selected email, approximately 10\% of characters were corrupted using random insertion, deletion, and substitution operations. 

\vspace{2pt}
\subsection{Robustness Evaluation Metric and Protocol}

To evaluate robustness under textual distortion, detection accuracy is used as the primary metric. Accuracy reflects the proportion of correctly detected emails under varying levels of character-level perturbation, providing a direct measure of model stability against realistic phishing evasion tactics.

\vspace{6pt}
For the test set, accuracy is defined as:
\vspace{6pt}
\begin{equation}
\text{Accuracy} = \frac{TP + TN}{TP + TN + FP + FN}
\end{equation}
\vspace{6pt}
where $TP$, $TN$, $FP$, and $FN$ denote true positives, true negatives, false positives, and false 
negatives, respectively.

\vspace{6pt}
Robustness is evaluated and compared across four DistilBERT variants under character-level perturbations.
\vspace{6pt}
\begin{itemize}
    \item \textbf{Baseline DistilBERT:} Trained on clean data without adversarial or noise augmentation.

    \item \textbf{FGM-DistilBERT:} Trained using embedding-level adversarial training with the FGM.

    \item \textbf{DistilBERT with character-level noise augmentation:} Trained using character-level 
    noise augmentation.

    \item \textbf{Proposed Framework:} Trained using both FGM-based embedding 
    perturbations and character-level noise augmentation.
\end{itemize}

\subsubsection{Robustness Evaluation Against Character-Level Perturbations}
To simulate realistic phishing obfuscation strategies, character-level perturbations are applied to the raw email text prior to tokenization. The evaluation assesses model resilience across increasing noise intensities: $0\%$, $5\%$, $10\%$, and $20\%$ corruption rates. For each level, identical perturbed samples are used across all models to ensure a controlled comparative analysis. The perturbation process stochastically selects from three primary operations:
\begin{itemize}
    \item \textbf{Deletion:} Random removal of characters.
    \item \textbf{Substitution:} Replacement of characters with visually similar symbols or random alphanumeric characters.
    \item \textbf{Insertion:} Addition of random alphanumeric characters at arbitrary positions.
\end{itemize}

\vspace{6pt}

\section{Results}\label{sec4}
\subsection{Robustness Analysis}

\FloatBarrier
\begin{table}[!ht]
\caption{Model Accuracy under Increasing Character-Level Noise (at Test Time)}
\label{tab:char_noise_accuracy}
\begin{tabular*}{\columnwidth}{@{\extracolsep\fill}lcccc}
\toprule
\textbf{Model Variant} & \textbf{0\% (Clean)} & \textbf{5\% Noise} & \textbf{10\% Noise} & \textbf{20\% Noise} \\
\midrule
DistilBERT Baseline            & 98.0 & 94.3 & 84.2 & 54.5 \\
DistilBERT + FGM               & 98.0 & 95.8 & 87.8 & 60.5 \\
DistilBERT + Char.-level Noise & 97.9 & 97.0 & 96.4 & 91.6 \\
Proposed Framework             & 98.1 & 97.3 & 97.0 & 93.2 \\
\botrule
\end{tabular*}
\end{table}
\FloatBarrier

Table 2 compares model accuracy under increasing levels of character-level noise applied at test time, providing a direct measure of robustness. The DistilBERT baseline exhibits a substantial degradation in performance as noise increases, with accuracy dropping from 98.0\% on clean data to 54.5\% at 20\% noise. This sharp decline highlights the baseline model's strong sensitivity to surface-level textual perturbations commonly used in phishing obfuscation.

FGM-based adversarial training offers moderate robustness improvements over the baseline while preserving comparable clean accuracy of 98.0\%. Accuracy improves to 60.5\% at 20\% noise, indicating that embedding-level perturbations enhance local robustness but remain insufficient to fully address character-level attacks. Training with explicit character-level noise results in significantly stronger robustness across all noise levels. This model achieves 96.4\% accuracy at 10\% noise and maintains 91.6\% accuracy at 20\% noise, demonstrating the effectiveness of exposure to realistic input corruptions during training.

The Proposed Framework delivers the best overall performance. It attains the highest clean accuracy of 98.1\% and shows superior stability under severe perturbations, achieving 93.2\% accuracy at 20\% noise. The total accuracy drop from clean data to 20\% noise is only 4.9 percentage points, compared to 43.5 percentage points for the baseline. These results indicate that combining FGM with character-level noise provides complementary robustness benefits. This combination enables the model to perform consistently well under both adversarial and noisy conditions. Overall, the findings demonstrate that the Proposed Framework Explainable and Robust DistilBERT significantly mitigates performance degradation caused by character-level attacks, making it well suited for reliable phishing detection in real-world adversarial environments.

\vspace{2pt}
\subsection{XAI Model Performance Comparison and Evaluation}
A qualitative robustness analysis is performed using an original email and a minimally perturbed version to simulate common phishing obfuscation strategies. Predictions and token-level attributions from LIME, SHAP, and IG are compared for the Baseline DistilBERT and proposed framework (Explainable and Robust DistilBERT model) to evaluate the stability of explanations under minor textual variations.

\vspace{6pt}
\textbf{Original Email:}
\textit{Our security team detected unauthorized actions linked to your account. Failure to confirm your identity within 30 minutes may result in legal escalation and service termination.}

\FloatBarrier
\begin{table*}[!ht]
\caption{Token-Level Feature Attributions by LIME, SHAP, and IG on the Original Email (Pre-Ablation)}
\label{tab:xai_original}
\begin{tabularx}{\textwidth}{>{\raggedright\arraybackslash}X >{\raggedright\arraybackslash}X >{\raggedright\arraybackslash}X >{\raggedright\arraybackslash}X}
\toprule
\textbf{Model \& Prediction (Conf.)} &
\textbf{LIME -- Top-10 Tokens} &
\textbf{SHAP -- Top-10 Tokens} &
\textbf{IG -- Top-10 Tokens} \\
\midrule
\textbf{Baseline DistilBERT} PHISHING (0.99) &
your (+0.0542) \newline
identity (+0.0352) \newline
unauthorized (+0.0280) \newline
to ($-$0.0269) \newline
confirm (+0.0172) \newline
result (+0.0166) \newline
actions ($-$0.0165) \newline
minutes (+0.0159) \newline
team ($-$0.0157) \newline
termination ($-$0.0121) &
your (+0.0273) \newline
your (+0.0263) \newline
minutes (+0.0225) \newline
account (+0.0224) \newline
unauthorized (+0.0213) \newline
security (+0.0152) \newline
service (+0.0131) \newline
es (+0.0111) \newline
legal (+0.0104) \newline
identity (+0.0101) &
your (+0.1448) \newline
failure (+0.1383) \newline
your (+0.1279) \newline
security (+0.1197) \newline
team (+0.1153) \newline
to (+0.1047) \newline
actions (+0.1040) \newline
account (+0.0797) \newline
linked ($-$0.0791) \newline
and (+0.0604) \\
\midrule
\textbf{Proposed Framework} PHISHING (0.99) &
your (+0.2754) \newline
account (+0.1009) \newline
team ($-$0.0973) \newline
unauthorized (+0.0822) \newline
to ($-$0.0797) \newline
identity (+0.0696) \newline
actions ($-$0.0613) \newline
may ($-$0.0554) \newline
termination ($-$0.0536) \newline
result (+0.0462) &
your (+0.0662) \newline
your (+0.0395) \newline
identity (+0.0363) \newline
account (+0.0341) \newline
may ($-$0.0250) \newline
result (+0.0194) \newline
to ($-$0.0158) \newline
legal (+0.0148) \newline
unauthorized (+0.0135) \newline
es (+0.0117) &
your (+0.8215) \newline
your (+0.6422) \newline
account (+0.5788) \newline
termination ($-$0.4729) \newline
service ($-$0.3478) \newline
failure (+0.3190) \newline
detected ($-$0.2338) \newline
identity (+0.1643) \newline
to (+0.1619) \newline
our (+0.1562) \\
\botrule
\end{tabularx}
\par\noindent\footnotesize $^*$ Positive scores increase phishing confidence; negative scores decrease it.
\end{table*}
\FloatBarrier

\textbf{Modified Email 1:}
\textit{Our team detected unauthorized actions linked to your acc0unt. Failure to confirm your identity within 30 minutes may reslt in legal escalation and service termmination.}

\vspace{4pt}

To evaluate explanation robustness, character-level perturbations and strategic keyword deletions were applied to the original email text, for instance, by omitting \textit{``security''} and applying transformations such as \textit{account} $\rightarrow$ \textit{acc0unt}, \textit{result} $\rightarrow$ \textit{reslt}, and \textit{termination} $\rightarrow$ \textit{termmination}. While preserving the overall semantic intent, these modifications reflect common phishing obfuscation tactics, enabling an assessment of explanation stability under realistic textual distortions.

\FloatBarrier
\begin{table*}[!ht]
\caption{Token-Level Feature Attributions by LIME, SHAP, and IG on Modified Email 1}
\label{tab:xai_modified_1}
\begin{tabularx}{\textwidth}{>{\raggedright\arraybackslash}X >{\raggedright\arraybackslash}X >{\raggedright\arraybackslash}X >{\raggedright\arraybackslash}X}
\toprule
\textbf{Model \& Prediction (Conf.)} &
\textbf{LIME -- Top-10 Tokens} &
\textbf{SHAP -- Top-10 Tokens} &
\textbf{IG -- Top-10 Tokens} \\
\midrule
\textbf{Baseline DistilBERT} PHISHING (0.96) &
your (+0.4114) \newline
identity (+0.1596) \newline
to ($-$0.1464) \newline
team ($-$0.1359) \newline
acc0unt ($-$0.1329) \newline
unauthorized (+0.1182) \newline
linked ($-$0.1007) \newline
minutes (+0.0976) \newline
reslt ($-$0.0942) \newline
legal (+0.0761) &
your (+0.0443) \newline
unauthorized (+0.0238) \newline
service (+0.0166) \newline
your (+0.0151) \newline
minutes (+0.0148) \newline
our (+0.0119) \newline
cala (+0.0102) \newline
actions (+0.0091) \newline
linked ($-$0.0078) \newline
acc (+0.0077) &
to (+0.5438) \newline
detected ($-$0.5371) \newline
and (+0.5282) \newline
your ($-$0.4934) \newline
res (+0.4072) \newline
failure ($-$0.3705) \newline
to (+0.3235) \newline
confirm ($-$0.3233) \newline
in (+0.3230) \newline
our ($-$0.3201) \\
\midrule
\textbf{Proposed Framework} PHISHING (0.98) &
your (+0.5359) \newline
team ($-$0.2523) \newline
unauthorized (+0.1011) \newline
to ($-$0.0843) \newline
acc0unt ($-$0.0776) \newline
linked ($-$0.0650) \newline
legal (+0.0513) \newline
identity (+0.0506) \newline
service (+0.0359) \newline
minutes (+0.0314) &
your (+0.0973) \newline
your (+0.0360) \newline
unauthorized (+0.0280) \newline
service (+0.0188) \newline
identity (+0.0158) \newline
in ($-$0.0151) \newline
to ($-$0.0139) \newline
linked ($-$0.0129) \newline
our (+0.0112) \newline
team ($-$0.0108) &
your (+0.8150) \newline
team ($-$0.7394) \newline
your (+0.6316) \newline
detected ($-$0.4873) \newline
our ($-$0.4165) \newline
identity (+0.3492) \newline
service ($-$0.3042) \newline
minutes ($-$0.2914) \newline
res (+0.2082) \newline
linked ($-$0.2076) \\
\botrule
\end{tabularx}
\par\noindent\footnotesize $^*$ Positive scores increase phishing confidence; negative scores decrease it.
\end{table*}
\FloatBarrier

\textbf{Modified Email 2:}
\textit{Our security team detected actions linked to your acount. Failure to conffirm your identity within 30 minutes may result in legal escalation and \$ervice termination.}

\vspace{2pt}
\noindent
To further stress explanation robustness, character-level perturbations and strategic token deletions were applied, for instance, by omitting the keyword \textit{``unauthorized''} and performing transformations such as \textit{account} $\rightarrow$ \textit{acount}, \textit{confirm} $\rightarrow$ \textit{conffirm}, and \textit{service} $\rightarrow$ \textit{\$ervice}.

\FloatBarrier
\begin{table*}[!ht]
\caption{Token-Level Feature Attributions by LIME, SHAP, and IG on Modified Email 2}
\label{tab:xai_modified_2}
\begin{tabularx}{\textwidth}{>{\raggedright\arraybackslash}X >{\raggedright\arraybackslash}X >{\raggedright\arraybackslash}X >{\raggedright\arraybackslash}X}
\toprule
\textbf{Model \& Prediction (Conf.)} &
\textbf{LIME -- Top-10 Tokens} &
\textbf{SHAP -- Top-10 Tokens} &
\textbf{IG -- Top-10 Tokens} \\
\midrule
\textbf{Baseline DistilBERT} PHISHING (0.98) &
your (+0.4877) \newline
acount ($-$0.2412) \newline
identity (+0.2126) \newline
actions ($-$0.1531) \newline
to ($-$0.1433) \newline
termination ($-$0.1305) \newline
linked ($-$0.0911) \newline
legal (+0.0904) \newline
security (+0.0862) \newline
team ($-$0.0681) &
your (+0.0706) \newline
security (+0.0376) \newline
team (+0.0190) \newline
detected ($-$0.0138) \newline
linked ($-$0.0125) \newline
our (+0.0120) \newline
to ($-$0.0104) \newline
termination (+0.0099) \newline
your (+0.0095) \newline
actions (+0.0083) &
to (+1.2027) \newline
ac ($-$0.8338) \newline
actions (+0.7732) \newline
detected ($-$0.7157) \newline
failure ($-$0.6368) \newline
and (+0.5421) \newline
to (+0.5117) \newline
identity (+0.5092) \newline
our ($-$0.5002) \newline
con (+0.4055) \\
\midrule
\textbf{Proposed Framework} PHISHING (0.98) &
your (+0.6886) \newline
team ($-$0.1738) \newline
actions ($-$0.1036) \newline
acount ($-$0.0766) \newline
may ($-$0.0731) \newline
identity (+0.0730) \newline
to ($-$0.0579) \newline
our (+0.0452) \newline
termination ($-$0.0405) \newline
security (+0.0365) &
your (+0.0758) \newline
your (+0.0243) \newline
to ($-$0.0182) \newline
identity (+0.0131) \newline
linked ($-$0.0123) \newline
security (+0.0111) \newline
our (+0.0092) \newline
team (+0.0078) \newline
termination (+0.0071) \newline
con (+0.0057) &
your (+1.0039) \newline
your (+0.7751) \newline
detected ($-$0.6599) \newline
identity (+0.4834) \newline
linked ($-$0.4644) \newline
team ($-$0.4173) \newline
termination ($-$0.2559) \newline
minutes ($-$0.1943) \newline
to (+0.1937) \newline
actions (+0.1397) \\
\botrule
\end{tabularx}
\par\noindent\footnotesize $^*$ Positive scores increase phishing confidence; negative scores decrease it.
\end{table*}
\FloatBarrier

The comparative analysis across the original email and two distinct modified variants demonstrates that the proposed Explainable and Robust DistilBERT framework exhibits substantially greater explanation stability under character-level perturbations compared to the Baseline DistilBERT model. For the original email, both models confidently classify the message as phishing and highlight semantically meaningful cues such as your, identity, unauthorized, account, and security across LIME, SHAP, and IG.

However, when character-level substitutions, insertions, deletions, and strategic keyword deletions are introduced in Modified Email 1, the Baseline model shows pronounced instability in its explanatory behavior. Although the prediction confidence remains high, token-level attributions shift markedly across all explanation methods. Several influential tokens from the original input are replaced by corrupted or fragmented forms or disappear entirely from the top-ranked explanations. IG explanations in particular exhibit large attribution swings toward syntactically weak tokens (e.g., to, and, in), indicating a brittle reliance on surface-level lexical patterns rather than semantic content.

This trend becomes more pronounced in Modified Email 2, where stochastic noise injection and token-level manipulations further amplify explanation instability in the Baseline model. Across LIME, SHAP, and IG, the Baseline explanations demonstrate substantial reordering of important tokens, polarity reversals, and attribution concentration on fragmented subword units (e.g., ac), reflecting heightened sensitivity to minor textual obfuscation commonly employed in phishing attacks.

In contrast, the proposed Explainable and Robust DistilBERT framework maintains comparatively stable attribution patterns across both modified emails. Despite noisy or corrupted tokens, core semantic indicators, including their perturbed variants, consistently remain influential across explanation methods. Attribution magnitudes change gradually rather than abruptly, and no key corrupted token disproportionately dominates the decision process. This behavior indicates that the hybrid training strategy that combines character-level noise injection with FGM-based adversarial training encourages reliance on more generalizable contextual representations rather than brittle surface forms. Overall, these results confirm that the proposed hybrid framework achieves more robust and interpretable phishing detection than the baseline DistilBERT model, beyond prediction confidence alone.

Some fragmented or unintuitive tokens observed in SHAP and IG explanations (e.g., ac, res) are attributable to WordPiece tokenization effects under character-level perturbation rather than failures of the explanation methods themselves. Additionally, repeated tokens reflect multiple subword occurrences resulting from WordPiece tokenization.

\vspace{2pt}
\subsection{Predictions with User-Centric Explanations}

Table 6 presents representative examples of phishing and legitimate emails, along with their model predictions and confidence scores, top LIME tokens, and plain-language explanations generated by the language model. This demonstrates the interpretability of the proposed approach on real-world email content.

\FloatBarrier
\begingroup
\footnotesize
\setlength{\tabcolsep}{4pt}
\renewcommand{\arraystretch}{1.15}
\captionsetup[longtable]{justification=raggedright, singlelinecheck=false}
\begin{longtable}{
>{\raggedright\arraybackslash}p{0.33\textwidth}
>{\raggedright\arraybackslash}p{0.14\textwidth}
>{\raggedright\arraybackslash}p{0.14\textwidth}
>{\raggedright\arraybackslash}p{0.31\textwidth}
}

\caption{Representative Examples of Model Predictions and Explanations on Real-World Emails}
\label{tab:real_world_examples} \\

\toprule
\textbf{Email Type \& Email Body} &
\textbf{Prediction (Confidence)} &
\textbf{Top LIME Tokens (Score)} &
\textbf{Language Model Explanation / Text Summary} \\
\midrule
\endfirsthead

\multicolumn{4}{c}{\tablename\ \thetable{} -- \textit{Continued from previous page}} \\
\toprule
\textbf{Email Type \& Email Body} &
\textbf{Prediction (Confidence)} &
\textbf{Top LIME Tokens (Score)} &
\textbf{Language Model Explanation / Text Summary} \\
\midrule
\endhead

\midrule
\multicolumn{4}{r}{\textit{Continued on next page}} \\
\endfoot

\bottomrule
\multicolumn{4}{l}{\footnotesize * (+) Boosts phishing confidence | (-) Lowers phishing confidence}
\endlastfoot

\textbf{Email 1: Phishing} Immediate action required! Your password has been compromised. Failure to act will result in permanent account deletion within 30 minutes. You must reset your password now to secure your account. Click the link below to verify. &
PHISHING (0.98) &
your (+) \newline
deletion ($-$) \newline
to ($-$) \newline
you (+) \newline
secure (+) \newline
account (+) \newline
failure (+) \newline
reset ($-$) \newline
password (+) \newline
immediate (+) &
The email was detected as PHISHING with high confidence (0.98). The message uses urgency and includes clickbait keywords, suggesting a fraudulent attempt to capture credentials or sensitive information. \\
\midrule

\textbf{Email 2: Legitimate} Please find attached the final agenda for the quarterly meeting tomorrow. I've made sure to include the notes we discussed. Also, please prepare a progress report presentation for the program. &
LEGITIMATE (0.96) &
report ($-$) \newline
agenda ($-$) \newline
discussed ($-$) \newline
meeting ($-$) \newline
final ($-$) \newline
include ($-$) \newline
presentation ($-$) \newline
also ($-$) \newline
quarterly ($-$) \newline
please (+) &
The email was detected as LEGITIMATE with high confidence (0.96). The message appears routine and contains no social-engineering cues or suspicious tokens. \\
\midrule

\textbf{Email 3: Phishing} We could not process your last payment. To prevent service interruption, reply to this email immediately with your full name and the last four digits of your credit card number. Service will be terminated in 50 minutes if we do not receive this information. &
PHISHING (0.99) &
your (+) \newline
payment (+) \newline
card (+) \newline
email (+) \newline
to ($-$) \newline
if ($-$) \newline
process ($-$) \newline
could ($-$) \newline
credit (+) \newline
service (+) &
The email was detected as PHISHING with high confidence (0.99). The message mentions sensitive financial terms and attempts to create a sense of financial risk or obligation. \\
\midrule

\textbf{Email 4: Legitimate} Thanks for speaking with our team yesterday. We will consolidate feedback and be in touch with an update on the next stage by Monday, November 1st. Best regards. &
LEGITIMATE (0.99) &
november ($-$) \newline
update ($-$) \newline
monday ($-$) \newline
team ($-$) \newline
regards ($-$) \newline
best (+) \newline
in ($-$) \newline
speaking ($-$) \newline
will ($-$) \newline
with ($-$) &
The email was detected as LEGITIMATE with high confidence (0.99). The message appears routine and contains no social-engineering cues or suspicious tokens. \\
\midrule

\textbf{Email 5: Legitimate (FP)} A new sign-in to your account was detected from an unknown device or location. If you do not recognize this activity, please secure your account immediately by opening our official app to review your recent activity and update your security settings. &
PHISHING (0.59) &
your (+) \newline
update (+) \newline
security (+) \newline
secure (+) \newline
official ($-$) \newline
app (+) \newline
activity ($-$) \newline
account (+) \newline
by ($-$) \newline
if ($-$) &
The email was detected as PHISHING with low confidence (0.59). The email uses urgency tactics suggesting a social-engineering attempt to capture sensitive information. \\
\midrule

\textbf{Email 6: Phishing (FN)} We noticed a minor issue during the processing of your recent subscription renewal. This occasionally occurs due to system updates. Please review the details in the attached document and let us know if any clarification is needed. &
LEGITIMATE (0.71) &
your (+) \newline
the ($-$) \newline
we ($-$) \newline
review ($-$) \newline
updates ($-$) \newline
details ($-$) \newline
issue ($-$) \newline
clarification ($-$) \newline
this ($-$) \newline
subscription (+) &
The email was detected as LEGITIMATE with moderate confidence (0.71). The message appears routine and contains no social-engineering cues or suspicious tokens. \\

\end{longtable}
\endgroup
\FloatBarrier

The results in Table 6 demonstrate the effectiveness of the proposed Framework in delivering both accurate predictions and user-centric interpretability for real-world email content. The language model translates the underlying feature attributions into concise, plain-language security explanations, enabling non-expert users to understand why an email is detected as phishing or legitimate. Correctly detected phishing emails (Emails 1 and 3) highlight the model's strong sensitivity to social-engineering cues, including urgency, threats of account compromise, financial risk, and explicit requests for sensitive information. Correspondingly, legitimate emails (Emails 2 and 4) are accurately identified due to their routine tone, contextual clarity, and absence of manipulation tactics, reinforcing the model's reliability in standard communication scenarios.

Importantly, the table also illustrates error cases that commonly occur in practical deployments. Email 5 represents a False Positive, where a legitimate security alert is incorrectly detected as phishing with low confidence (0.59). This misclassification is mainly driven by urgency cues and account-related warnings that resemble common phishing patterns, causing the model to prioritize threat-related language despite the message advising safe actions such as using the official app. 

Conversely, Email 6 demonstrates a False Negative, where a phishing attempt is incorrectly detected as legitimate with relatively moderate confidence (0.71). The message avoids explicit urgency, threats, or credential requests, instead adopting a neutral service-notification tone and using an attachment-based delivery strategy. This highlights a known limitation of keyword-driven and attention-based classifiers when faced with subtle, low-signal phishing emails. More importantly, the generated explanations reveal the model's reasoning behind these misclassifications, allowing users and security analysts to apply secondary verification strategies when confidence is minimal or when explanations reference borderline cues. This reinforces the framework's role not only as a detection system, but as a decision-support tool that enhances transparency, trust, and practical usability in email security contexts.

\vspace{2pt}
\subsection{Cross-Domain Dataset Validation}
To evaluate generalization, cross-dataset validation was conducted using the \texttt{CEAS\_08} subset of the Phish No More corpus \cite{alam2022kaggle_dataset}, which contains 39,154 emails with a moderate class imbalance (55.78\% suspicious and 44.22\% non-suspicious). The subset was divided into training (70\%), validation (15\%), and test (15\%) partitions using stratified sampling. Class-weighted cross-entropy loss was applied during training to compensate for class imbalance, with weights computed from the training set.

This dataset is mixed and primarily spam-oriented, comprising spam, ham (legitimate), and phishing-related emails, with each record including raw email text and a corresponding label for binary classification. Although such datasets are commonly used for spam detection, they are also suitable for phishing detection research because spam and phishing emails often share overlapping linguistic characteristics, such as persuasive language, urgency cues, and deceptive intent. Accordingly, spam and phishing-related samples were mapped to the suspicious class (label = 1), while ham and legitimate emails were mapped to the non-suspicious class (label = 0) for consistency with the primary dataset. Cross-dataset evaluation was performed under the same character-level noise conditions used in the in-domain robustness experiments to assess robustness to input distortions across domains. Performance is reported for two trained models: The Baseline DistilBERT and the Proposed Framework (Explainable and Robust DistilBERT), using identical data splits and evaluation settings.

\vspace{4pt}

\textbf{Cross-Domain Dataset Evaluation on Kaggle Phish No More: CEAS Dataset}

\FloatBarrier
\begin{table}[!ht]
\caption{Cross-Domain Model Accuracy under Increasing Character-Level Noise (at Test Time)}
\label{tab:ceas_char_noise_accuracy}
\begin{tabular*}{\columnwidth}{@{\extracolsep\fill}lcccc}
\toprule
\textbf{Model Variant} & \textbf{0\% (Clean)} & \textbf{5\% Noise} & \textbf{10\% Noise} & \textbf{20\% Noise} \\
\midrule
DistilBERT Baseline & 99.4 & 98.3 & 92.4 & 67.6 \\
Proposed Framework  & 99.4 & 99.1 & 98.7 & 95.7 \\
\botrule
\end{tabular*}
\end{table}
\FloatBarrier

Table 7 evaluates cross-domain robustness on the CEAS dataset under increasing levels of character-level noise, simulating realistic obfuscation strategies commonly observed in phishing emails. The DistilBERT baseline achieves near-perfect performance on clean inputs (99.4\%); however, its accuracy degrades sharply as noise intensity increases, falling to 92.4\% at 10\% corruption and further to 67.6\% at 20\% noise. This steep performance decline highlights the baseline model's limited robustness to character-level perturbations when trained without explicit noise-aware or adversarial defenses. 

In contrast, the proposed framework, which integrates FGM-based adversarial training with stochastic character-level noise injection, maintains consistently high accuracy across all noise conditions. Performance remains consistent at low and moderate noise levels (99.1\% at 5\% and 98.7\% at 10\%), and degrades only marginally under severe corruption (95.7\% at 20\% noise). These results demonstrate that joint exposure to embedding-level adversarial perturbations and character-level input noise during training significantly enhances cross-domain generalization and robustness to unseen, corrupted inputs.

\subsection{User Study: Pilot Evaluation of Practical Usefulness and Explanation Clarity}
To conduct an exploratory pilot user study of the proposed robust explainable phishing classification framework, a web-based interactive prototype was developed and deployed. The primary objective was to assess practical applicability, system usability, and user perceptions of explanation quality, as well as to validate the feasibility of the evaluation protocol rather than to perform a benchmark comparison or draw definitive conclusions. The system supports real-time inference, explanation delivery, and structured user feedback collection.
Participants were recruited through voluntary invitations shared on social media platforms, with no restrictions on background or technical expertise. A total of 30 participants took part in the pilot study. The study was not intended to provide statistically generalizable conclusions, but to identify early usability trends and to assess whether the explanations are perceived as understandable and informative in realistic usage conditions.

Users could enter any email text of their choice. For each submission, the system generated a predicted class label (phishing or legitimate) and a confidence score. High-impact tokens identified using LIME-based feature attribution were further summarized into a short natural-language explanation by the explanation module. The backend inference service was implemented using FastAPI, and the application was deployed on Hugging Face Spaces. User feedback was securely stored using Google Firebase to support structured data collection.

After reviewing the outputs, participants completed a short 5-point Likert-scale questionnaire (1 = Strongly Disagree, 5 = Strongly Agree). The questionnaire evaluated two dimensions: Decision Clarity, measuring how clearly the explanation conveyed the rationale behind the prediction, and Information Focus, assessing whether the explanation emphasized relevant phishing indicators rather than generic or irrelevant content. These measures reflect both interpretability and practical decision support, which are critical for real-world phishing detection tools.

\vspace{6pt}
\textbf{Usefulness and Clarity of Explanations Results Based on User Feedback}

\FloatBarrier
\begin{table}[!ht]
\caption{User Feedback on Explanation Quality ($N = 30$)}
\label{tab:user_feedback}
\begin{tabular*}{\columnwidth}{@{\extracolsep\fill}llcc}
\toprule
\textbf{Metric} & \textbf{Statement} & \textbf{Mean ($\mu$)} & \textbf{Std.\ Dev.\ ($\sigma$)} \\
\midrule
Decision Clarity  & The explanation clarified the result.  & 4.34 & 0.55 \\
Information Focus & The explanation was concise and essential.        & 4.22 & 0.73 \\
\botrule
\end{tabular*}
\end{table}
\FloatBarrier

Table 8 summarizes the aggregated user feedback across two evaluation dimensions: Decision Clarity and Information Focus. The high mean score for Decision Clarity ($\mu = 4.34$) indicates that participants generally found the explanations effective in conveying the rationale behind the phishing predictions. This suggests that emphasizing salient cues, such as urgency indicators, account-related terms, and suspicious actions, aligns well with users' intuitive understanding of phishing behavior. The relatively low standard deviation ($\sigma = 0.55$) further reflects strong consistency in participant responses.

For Information Focus, the mean score ($\mu = 4.22$) indicates that users perceived the explanations as concise and centered on essential information. Although the standard deviation is slightly higher ($\sigma = 0.73$), suggesting greater variability in user perceptions compared to Decision Clarity, the overall high mean still reflects a positive evaluation. This variability suggests that while most users found the explanations appropriately focused, some perceived room for improvement in streamlining or prioritizing key details. This highlights an opportunity to refine explanation brevity without sacrificing interpretability.

\section{Discussion}\label{sec5}
\textbf{Impact of Adversarial Training on Model Robustness}

The results demonstrate that embedding-level adversarial training using FGM marginally improves the robustness of transformer-based phishing classifiers under both moderate and severe perturbations. This indicates that FGM promotes more stable and generalizable linguistic representations while maintaining comparable clean-data accuracy. These findings highlight the practicality of adversarial defenses for real-world email filtering.

\vspace{6pt}
\textbf{Impact of FGM-Based Adversarial Training and Character-Level Perturbation on Model Robustness}

The findings indicate that FGM improves stability in the representation space, thereby supporting stronger generalization to unseen perturbations and cross-dataset settings. Character-level noise augmentation further enhances robustness by exposing the model to realistic textual obfuscation that affects tokenization and lexical structure. The combined strategy substantially reduces performance degradation under severe corruption while preserving clean-data accuracy. These results suggest that embedding-level adversarial robustness can be better aligned with surface-level textual variations when reinforced through character-level augmentation, validating the hybrid approach for deployment in adversarial phishing detection environments.

\vspace{6pt}
\textbf{Impact of Robustness-Enhancing Training on Explanation Stability}

Beyond predictive robustness, the proposed hybrid robustness-enhancing training strategy also improves the stability and coherence of model explanations. Embedding-level adversarial perturbations encourage smoother decision boundaries in the representation space, leading to more consistent gradient-based and perturbation-based token attributions under small input changes. In addition, character-level noise augmentation reduces sensitivity to superficial spelling variations and tokenization artifacts, encouraging explanations to rely more on semantically meaningful patterns rather than surface-form cues. Token ablation analysis shows that the robustly trained model redistributes importance to related semantic tokens instead of exhibiting abrupt confidence shifts, resulting in more reliable post-hoc explanations and more coherent natural-language rationalizations.

\vspace{6pt}
\textbf{Robustness Under Test-Time Character-Level Noise}

Character-level noise is applied at test time to assess robustness under realistic phishing obfuscation, even though FGM and character-level noise are used during training. Real-world attacks often introduce unseen and more severe perturbations that differ from training distortions. Test-time noise therefore acts as a stress test, ensuring that performance gains reflect true generalization rather than adaptation to known perturbations. This evaluation verifies that the Proposed Framework remains stable under adversarial and noisy deployment conditions.

\vspace{6pt}
\textbf{Comparative Behavior of XAI Methods}

A comparative analysis of LIME, SHAP, and IG reveals complementary explanatory behaviors. LIME and SHAP often produce relatively sparse, human-interpretable attributions that strongly align with intuitive phishing indicators, such as urgency-driven language, suspicious call-to-action prompts, or domain-related cues. In contrast, IG, a gradient-based method, exhibits higher sensitivity to the model's contextual dependencies and long-range interactions captured by attention mechanisms. It tends to distribute attribution across a broader set of influential tokens (including semantically related words and sometimes contextual fillers), resulting in more distributed importance scores while still assigning higher-magnitude scores to the most critical tokens that reflect the model's internal mathematical sensitivity.

\vspace{6pt}
\textbf{Failure Case Analysis and Error Patterns}

Analysis of the system’s performance reveals two primary error patterns. False Negatives (FN) typically occur with highly refined spear phishing or AI-generated messages that closely mimic benign language and avoid obvious trigger terms, thereby allowing malicious content to evade detection. In contrast, False Positives (FP) occur when legitimate urgent communications, such as authentic alerts or financial notifications, contain high urgency language and are therefore misclassified as phishing.

\subsection{Limitations and Challenges}

\begin{itemize}
    \item Although BERT-style models improve text detection accuracy, they remain computationally expensive due to the large number of parameters in full-scale BERT-style models. Training and deployment on large-scale, real-time email streams may require substantial hardware resources.

    \item Cyber adversaries continuously evolve, requiring detection systems to be periodically updated and retrained. While character-level perturbations and embedding-level adversarial training improve robustness to common obfuscation tactics, they do not fully capture optimized adversarial attacks, semantic paraphrasing, or socially engineered linguistic strategies, which may limit generalization to more sophisticated phishing campaigns.

    \item Although plain-language explanations enhance interpretability, they may introduce ambiguity, linguistic bias, and limited contextual understanding, potentially affecting clarity and user trust. Moreover, XAI methods such as LIME, SHAP, and IG provide approximate, attribution-based explanations rather than direct insights into internal decision processes.

    \item The proposed system relies on publicly available datasets, which supports reproducibility but may not fully reflect the diversity and sophistication of real-world phishing attacks. In addition, the dataset is limited to English text and does not include multilingual or image-based phishing scenarios.
\end{itemize}

\subsection{Ethical Considerations}

Ethical responsibility is critical in the cybersecurity domain, particularly in the case of phishing, where sensitive user data is usually involved. This study utilizes publicly available phishing email datasets to ensure transparency and reproducibility. While the proposed adversarial training enhances defensive robustness, it raises the dual-use risk. Adversaries could apply similar hardening techniques to malicious LLMs, enabling them to generate more evasive phishing content that bypasses detection systems. The adversarial attacking method FGM used in this study is strictly for defensive purposes to enhance resilience against malicious phishing attacks and must not be repurposed for harmful activities. 

The explainability mechanism is designed to enhance transparency and user trust by providing clear reasoning for why an email is detected as phishing or legitimate. However, these explanations are intended to support user understanding rather than replace human judgment, and the final assessment should always be verified by the end user for accuracy and reliability.
The user study was exploratory in nature, and participation was entirely voluntary. Participants were recruited through open invitations shared on social media platforms, including a brief study description and a link to the web-based system. No identifying information was collected, and participants remained fully anonymous; the researcher had no means of knowing who participated. No personal details or email contents were recorded, and only Likert scale ratings related to explanation quality were collected from the first 30 participants.

\subsection{Reproducibility}
To ensure full reproducibility of the proposed framework and experimental results, all essential resources and configurations are made publicly available. Specifically:

\vspace{6pt}
\textbf{Code:} The complete implementation of the model is provided in a public GitHub repository \cite{sajstack2024github_repo}.

\vspace{6pt}
\textbf{Datasets:} All datasets used in this study are publicly accessible and were obtained from Hugging Face \cite{liu2024phishing_dataset} and Kaggle \cite{alam2022kaggle_dataset}.

    \vspace{6pt}
\textbf{Environment:} Experiments were conducted using Python 3.12.12, PyTorch 2.9.0, and Hugging Face Transformers 4.57.6.

\vspace{6pt}
\textbf{Hardware:} Model training and evaluation were performed on Google Colaboratory with an allocated GPU (e.g., NVIDIA Tesla T4 with 16 GB VRAM).

\section{Conclusion}\label{sec6}

This study presented a unified, lightweight, and interpretable phishing detection framework that integrates robustness-enhancing training, XAI, and attribution-grounded plain-language explanations. By applying FGM-based embedding-level adversarial perturbations, the proposed DistilBERT model improves resilience to adversarial inputs while maintaining strong clean-data detection performance. In addition, character-level noise augmentation is incorporated during training to enhance robustness against realistic text obfuscation attacks, providing an efficient alternative to more computationally expensive defense strategies.

A systematic comparison of LIME, SHAP, and IG using character-level perturbations and selective token removal demonstrated the robustness of these attribution methods when applied to robustly trained transformer models. The results also highlighted their complementary behaviors, providing insight into the interpretability of the model under robustness constraints. Furthermore, a simplified explanation layer based on Flan-T5-Small was introduced to translate feature attributions and predictive insights into concise, user-friendly security narratives, enhancing transparency and user trust. Beyond phishing detection, the design principles proposed in this work offer a scalable and transparent foundation that can be extended to broader cybersecurity threat detection and classification tasks in future research.

\subsection{Future Work}
\begin{itemize}
    \item Incorporating continuous adaptive learning via an adversarial training loop to maintain long-term robustness in real-world deployments, allowing the model to evolve against sophisticated linguistic and structural attack strategies.

    \item Applying federated learning to enable collaborative training that preserves privacy within organizations without sharing sensitive data, thus improving scalability and generalization.

    \item Extending the proposed framework to other high-priority domains such as healthcare cybersecurity, critical infrastructure protection, financial systems, and Internet of Things (IoT) security.

    \item Exploring paraphrasing and back-translation-based adversarial example generation to enhance robustness against linguistic manipulation, along with extensions to multilingual and multimodal phishing detection, including image- and QR-code-based attacks.
\end{itemize}

\section*{Conflict of Interest Statement}
No competing financial or personal interests influenced this work. No funding was received.

\section*{Acknowledgements}
The author expresses appreciation to the research community and dataset developers, and extends gratitude to Hugging Face for hosting the Flan-T5-Small model. The design of this method was informed by previous studies on transformer-based phishing classification, adversarial robustness, and XAI. The Python prototype implementation benefited from the use of AI language models as coding aids, and AI-based language tools were also used to improve grammar, clarity, and paraphrasing in the manuscript. AI tools were not used to generate original research ideas, perform experiments, analyze datasets, or draw conclusions. All results and interpretations were produced and verified by the author.


\bibliography{sn-bibliography}

\end{document}